\documentclass[10pt,twocolumn,
amssymb, nobibnotes, notitlepage,nofootinbib aps, prd]{revtex4}

\usepackage{amsmath}
\usepackage{amssymb}
\usepackage{amsthm}
\usepackage[pdftex]{color}
\usepackage[pdftex,colorlinks,citecolor=blue,linkcolor=blue,urlcolor=blue]{hyperref} 
\usepackage{graphicx}
\usepackage{dcolumn} 
\usepackage{bm} 

\usepackage{longtable}

\usepackage{ulem}   
\usepackage{comment} 
\usepackage{datetime}

\usepackage{tabularx}
\usepackage{float}
\restylefloat{table}

\newcommand{\beq}{\begin{equation}}
\newcommand{\eeq}{\end{equation}}
\newcommand{\bea}{\begin{eqnarray}}
\newcommand{\eea}{\end{eqnarray}}
\newcommand{\barr}{\begin{array}}
\newcommand{\earr}{\end{array}}

\long\def\begincomment#1\endcomment{}

\newcommand{\mbZ}{\mathbb{Z}}

\newtheorem{remark}{Remark}

\begin{document}


\title{ Spherically symmetric  potential   in 
 noncommutative spacetime with a
 compactified 
extra dimensions
}

\author{S\^ecloka Lazare Guedezounme} \email{guesel10@yahoo.fr}  
\affiliation{International Chair in Mathematical Physics and Applications (ICMPA-UNESCO Chair), University of Abomey-Calavi,
072B.P.50, Cotonou, Republic of Benin}

\author{Antonin Danvid\'e Kanfon} 
\email{kanfon@yahoo.fr}
\affiliation{Facult\'e des Sciences et Techniques,  University of
d'Abomey-Calavi, Benin}
\affiliation{International Chair in Mathematical Physics and Applications (ICMPA-UNESCO Chair), University of Abomey-Calavi,
072B.P.50, Cotonou, Republic of Benin}

\author{Dine Ousmane Samary}
\email{dine.ousmane.samary@aei.mpg.de}

 \affiliation{Max Planck Institute for Gravitational Physics, Albert Einstein Institute, Am M\"uhlenberg 1, 14476, Potsdam, Germany}
\affiliation{Facult\'e des Sciences et Techniques,  University of
d'Abomey-Calavi, Benin}
\affiliation{International Chair in Mathematical Physics and Applications (ICMPA-UNESCO Chair), University of Abomey-Calavi,
072B.P.50, Cotonou, Republic of Benin}

\date{\today}

\begin{abstract}The Schr\"odinger equation of the spherical symmetry quantum models such as the hydrogen atom problem seems to be analytically non-solvable in higher dimensions. When we try to compactifying one  or several dimensions this question can maybe solved. This paper stands for the study of  the spherical symmetry quantum models on noncommutative spacetime with compactified  extra dimensions. We provide analytically the resulting spectrum of the hydrogen atom and Yukawa problem in $4+1$ dimensional noncommutative spacetime in the first order approximation of noncommutative parameter.  The case of higher dimensions $D\geq 4$ is  also discussed.
\end{abstract}


\maketitle

\section{Introduction} 
One of the recently discovered concepts that has impacted the theoretical
physics community in a major way is most likely the idea of a
noncommutative (NC) spacetime which led to a NC generalization
of quantum mechanics and field theory. The idea of noncommutativity of spacetime was first discussed in the work by Snyder
\cite{Snyder:1946qz} and  Connes \cite{Connes:1994yd}-\cite{Connes:1990qp}. The above concept (NC) spacetime allows to find  possible solution
to ultraviolet divergencies in quantum field theory \cite{Szabo:2001kg}-\cite{Grosse:2004yu}. 
The NC physics also arises as a possible scenario for the short-distance
behaviour of physical theories (the Planck scale). At this  scale, 
 the universal constants  $c, \hbar$ and $G$ appear naturally equivalent.  Under the Planck length, the
distance loses its meaning \cite{Szabo:2001kg}-\cite{Doplicher:1994tu} and the physical phenomena are believed to be nonlocal. NC geometry could be realized by introducing the noncommutativity through the coordinates which satisfy the commutation relations $[ x^\mu, x^\nu]=i\theta^{\mu\nu}$, where $\theta^{\mu\nu}$ is a skew-symmetric matrix characterizing the deformation of the spacetime. 
This leads to a new  Heisenberg uncertainty relation, given on the spacetime coordinates by $\Delta x^\mu\Delta x^\nu\geq \theta^{\mu\nu}$, and makes this spacetime a quantum space \cite{Doplicher:1994tu}-\cite{Majid:1999tc}. The important implications of noncommutativity is the loss of Lorentz invariant in the dispersion relations  and the loss of causality \cite{Ferrari:2005ng}-\cite{Berrino:2002ss}. 
 Intuitive arguments involving quantum mechanics in NC space is realized by imposing the commutation relations, now  between coordinates  and momentums as 
\bea
[ x^\mu, x^\nu]=i\theta^{\mu\nu},\quad [p^\mu,p^\nu]=i\gamma^{\mu\nu},\quad  [x^\mu, p^\nu]=i\hbar \kappa^{\mu\nu}
\eea
 where $\gamma^{\mu\nu}$  is also skew symmetric matices. In this paper we restrict ourself to the case where $\gamma^{\mu\nu}=0$, this implies that $\kappa^{\mu\nu}=\delta^{\mu\nu}$, the Kronecker symbole. We also assume that the tensor $\theta^{\mu\nu}$ is chosen to have the dimension of $length\cdot time$ i.e. $\theta^{0j}=\theta^j\in\mathbb{R}, \,\,\theta^{ij}=0,\,\,
 \,\,i, j=1,2,\cdots, D.$
The noncommutative  variables can be expressed in terms of commutative coordinates as $x^j=x_c^j-i\theta^j\partial_0=( x ),$ and $ p^j=p^j_c,\,\, p^0=i\hbar\partial_0=E$, where  the index ``c''  is used to specify the commutative variables and where $E$ is the energy of the system. The Hamiltonian of quantum system on NC space can be expressed with the commutative coordinates $H(x,p)\equiv H_c(x_c,p_c,\theta)$, where the parameter $\theta=\theta^j$ is showed to have the fundamental limit $\theta\preceq 1.6\cdot 10^{-27}m\cdot s\approx 0.3 (keV)^{-2}$ which is smaller than the one obtained by the theory of quantum gravity   \cite{Moumni:2010yf}-\cite{Moumni:2009dc}. 

 The compactified extra dimension is motivated by string theory, which predicts the existence of extra dimensions and   noncommutativity beetween coordinates. Our idea is to understand how the eigenvalue problem changes if we periodically identify one of the NC coordinates $x^j=(x^1,x^2,x^3,x^4)$ in the target space, say  $[-\pi R,\pi R]\ni w, \text{such that}\,\, x^4=w - 2\pi R k,\,\, k\in\mathbb{Z}$ and $R$ is the radius of the circle.  The wave function $\psi(x^0,x^{\bar\ell},x^4),\,\bar\ell=1,2,3$, can be expanded in the Fourier mode as \cite{Elze:2003tb}
\bea
\psi(x^0,x^{\bar\ell},x^4)=\frac{1}{\sqrt{2\pi R}}\sum_{n=-\infty}^\infty\, \psi_n(x^0,x^{\bar\ell})\exp\big( \frac{i n\, x^4}{R}\big).
\eea
Note that  the orthonormalized functions $(2\pi R)^{-1/2}\exp\big(i x^4 n/R\big)$ are eigenfunctions of the operator $\nabla^2_{x^4}$ with eigenvalues $E_{n0}=-n^2/R^2$. This means that the spectrum of quantum systhem defined with one dimension compactify is in the form $E_{nl}=E_{n0}+E'_{nl}$, where $l$ is a positive integer, $E'_{nl}$ depends on the potential associated to the system which requires to be computed.
In the following paper we investigate   the spectrum of the Coulomb and Yukawa Hamiltonian  on $(4+1)$-dimensional NC spacetime. Using the first order approximation of the deformation parameter  $\theta$ and by compactifying one extra dimension  $x^4$ resulting topology $\mathbb{R}^{3+1}\times S^1$ (see \cite{Bures:2015hfa} and \cite{Bures:2014fza} for the essential reviews), the spectrum  may be given exactly. 
We prove that in the case of ``$space\cdot time$ '' noncommutativity, the correction of the energy spectrum  does not depend on the NC deformation parameter $\theta$ but rather on the dimension compactified parameter.   

 Our paper is organized as follows. In section $2$, we focus on the hydrogen atom in $(D+1)$-dimensional noncommutative space with non-compactified extra dimension. We discuss the particular case where $D= 4$ in which the solution of the spectral problem can be solved. The Yukawa potential is also discussed in this section. In section $3$ the same problem is solved with now compactified extra dimensions. The discussion and conclusion are given in section $4$.

\section{Hydrogen atom in noncommutative space with non-compactified extra dimension}\label{sec2}
In this section we focus on the hydrogen atom problem defined in $(D+1)$ dimensional  NC spacetime (we consider the particular case where $D=4$). To be specific, the model is given with  the spherical potential of the form
\bea\label{potent}
V(\vec r_{nc})=-\frac{q_e^2}{|\vec r_{nc}|^{D-2}}, 
\eea
where $q_e$ is related to the atomic charge and where we  use the following notation $\vec{r}_{nc}=\vec{r}-i\vec\theta\partial_0$, i.e, (the NC coordinates are $\vec r_{nc}=(x)$) and (the commutative coordinates are $\vec r=(x_c)$).   It would be advisable to work in spherical coordinates system $\vec r=(r,\alpha_1,\alpha_2,\alpha_3)$ such that $r\in\mathbb{R}_+$, $0<\alpha_{\bar\ell}<\pi,\,\, \bar\ell=1,2$ and $0<\alpha_{3}<2\pi$. It thus follows that the Hamiltonian of the system is
\bea\label{Hamilton}
H=-\frac{\hbar^2}{2m}\Big[\frac{\partial^2}{\partial r^2} + \frac{D-1}{r}\frac{\partial}{\partial r}-\frac{\mathcal{L}^2(D-1)}{r^2}\Big]+V(\vec{r}_{nc}),
\eea
where $\mathcal{L}^2(D-1)$ is the Laplace-Beltrami operator on the $(D-1)$-sphere. Hence 
the potential \eqref{potent}, using  the first order Taylor expansion on $\vec\theta$ is
\bea\label{intpot}
V(\vec r_{nc})&=&-\frac{ q_e^2}{|\vec r-i\vec\theta\partial_0|^{D-2}}\cr
&\approx&  - \frac{q_e^2}{r^{D-2}}\Big(1 + i(D-2)\frac{\vec{r}.\vec{\theta}}{r^2}\partial_0 \Big).
\eea
We consider the adequate choice, such that the vector $\vec\theta$ is transform in the spherical coordinates as  $\vec{r}.\vec{\theta}\equiv  r \theta$ \cite{Moumni:2010yf}. Furthermore, the  spherical function $ \mathcal{Y}_\ell^{(D-1)}(\alpha_1,\alpha_2, \cdots, \alpha_{D-1})$, which are the eigenfunctions of  the operator $\mathcal L(D-1)$ is considered:
\begin{eqnarray}\label{harmonic}
\mathcal{L}^2 (D-1) \mathcal{Y}_\ell^{(D-1)} = \ell(\ell+D-2) \mathcal{Y}_\ell^{(D-1)} = \lambda_D \mathcal{Y}_\ell^{(D-1)},
\end{eqnarray}
 where $\ell$ is the orbital angular momentum quantum number.

Note that the Hamiltonian \eqref{Hamilton} depend on the partial derivative  with respect to the time $t$, due to the relation \eqref{intpot}. But
one can show that, the wave function $\psi(\vec r,t)$, namely the solution of the Schr\"odinger equation is expressed as $\psi (\vec{r},t) = \mathcal{Y}_\ell^{(D-1)}\psi (r)f(t)$,  where the time dependent function is
$
f(t) = \exp\big({-\frac{i}{\hbar}Et}\big)
$
and where $\psi(r)$ satisfied the radial equation 
\begin{eqnarray} \label{sch0}
&\Big[\frac{d^2 }{dr^2} + \frac{D-1}{r}\frac{d}{dr}-\Big(\frac{\lambda_D}{r^{2}} - \frac{\nu^2}{r^{D-2}} - \frac{\mu_D}{r^{D-1}} - \alpha^2\Big)\Big]\psi(r) \cr
&= 0,
\end{eqnarray}
with
$
\alpha^2 = \frac{2m E}{\hbar^2},\,\, \nu^2 = \frac{2m q_e^2}{\hbar^2} ,\,\,\mu_D = (D-2) \nu^2 \theta\, E/\hbar.
$
Now for  $D=4$, this equation  turns to be:
\begin{eqnarray} \label{sch1}
\Big[\frac{d^2 }{dr^2} + \frac{3}{r}\frac{d}{dr}-\Big(\frac{\lambda_4-\nu^2}{r^{2}}  - \frac{\mu_4}{r^{3}} - \alpha^2\Big)\Big]\psi(r) = 0.
\end{eqnarray}
However in this case, (unlike for the commutative case discussed in \cite{Bures:2015hfa}-\cite{Bures:2014fza}), such equation seems to be non-solvable.
We provide an algebraic method, which will allow us to derive the solution of this equation. For this, let us reparameterized the function $\psi(r)$ as
$\psi(r):=\psi( r, \theta\,)$.  Then $\psi(r,0)$ corresponds to the solution of the equation \eqref{sch1} in the case where $\theta=0$. The first order Taylor  expansion  on $\theta$ of the function  $\psi(r,\theta)$ takes the form
\begin{eqnarray}\label{sol1}
\psi(r,\theta\,) &=& \psi(r,0) + \theta\, \frac{d\psi(r,\theta\,)}{d\theta\,}\Big|_{\theta\, = 0} + \mathcal{O}(\theta\,^2).
\end{eqnarray} 
We get simply
\begin{eqnarray}
 \psi(r,0) = \frac{{\it c}}{r}{ J}_{\nu}(\alpha r)+
\frac {{\it c'}}{r}{Y}_{\nu}( \alpha r),\quad c, \,c'\in\mathbb{R},
\end{eqnarray}
where ${ J}_{\nu}(\alpha r)$ and ${Y}_{\nu}(\alpha r)$ are respectively the first  and second  kind  Bessel functions (see \cite{Bures:2014fza} for more detail). By replacing the solution \eqref{sol1} in the partial differential equation \eqref{sch1}, we get
\begin{eqnarray}\label{eqnow}
 \Big[\frac{d^2 }{dr^2} + \frac{3}{r}\frac{d}{dr}
+ \Big(\alpha^2 - \frac{\lambda_4-\nu^2}{r^{2}}\Big)\Big]\tilde{\chi}(r) = -\frac{2\nu^2E}{\hbar r^3}\psi(r,0), 
\end{eqnarray}
where $ \tilde{\chi}(r) = \frac{d\psi(r,\theta\,)}{d\theta\,}\Big|_{\theta\, = 0}$.
This equation corresponds to a nonhomogeneous differential equation, which can be solved easly.  For $\epsilon=(1+\lambda_4-\nu^2)^{1/2}$ and $g(r)=-2\nu^2E\psi(r,0)/\hbar r^3$, by using the  Wronskian method, the solution of the equation \eqref{eqnow} takes the form
\bea\label{malade}
\tilde{\chi}(r)&=&cJ_\epsilon(\alpha r)+c'Y_\epsilon(\alpha r)\cr
&-& \frac{\pi}{2r}J_\epsilon(\alpha r)\int_1^r x^2 Y_\epsilon(\alpha x) g(x)dx\cr
&+&\frac{\pi}{2r}Y_\epsilon(\alpha r)\int_1^r x^2 J_\epsilon(\alpha x) g(x)dx,\; c, \,c'\in\mathbb{R}.\;
\eea
Remark that there are some difficulties however.  One defect of this method  (in the commutative and NC case) is that the energy spectrum  can only be determined numerically, and we do not deal here with a numerical method to provide this  spectrum. In more than $(4+1)$ dimensions,  the differential equations  \eqref{sch0} are much more complicated to be solve.

Now let us  discuss the case of Yukawa potential :
\begin{eqnarray}\label{Yukawa}
V(\vec r_{nc}) = -V_0 \frac{e^{-\eta r_{nc}}}{|\vec r_{nc}|^{D-2}},
\end{eqnarray}
where $V_0 $ and  $\eta$  depend on the  constant of the neutral atom. In order to probe this potential, we write the expression \eqref{Yukawa} at the first order on  $\theta$ as
\begin{eqnarray}
V(\vec r_{nc},t)= -\frac{V_0 e^{-\eta r}}{r^{D-2}}\Big[1+i\big(\eta  r+D-2\big)\frac{\theta}{r}\partial_0 \Big].
\end{eqnarray}
 After separation variables  in the Schr\"odinger equation,
 it become easy to show that  the radial equation is given by the following:
\begin{eqnarray}
&&\Big[\frac{\partial^2}{\partial r^2} + \frac{D-1}{r}\frac{\partial}{\partial r}+s_D(r)\Big]\psi (r) = 0.
\end{eqnarray}
where 
\bea
s_D(r)&=&
 \frac{2m V_0 e^{-\eta r}}{\hbar^3 r^{D-2}}\Big(D-2+ \eta r \Big)\frac{E\theta }{r} \cr
&+&  \frac{2m E}{\hbar^2} -\frac{\lambda_D}{r^2} +  \frac{2m V_0 e^{-\eta r}}{\hbar^2 r^{D-2}} ,\nonumber
\eea 
In the particular case where $D = 4$,  this equation is reduced to
\begin{eqnarray}
&&\frac{d^2 \psi (r)}{d r^2} + \frac{3}{r}\frac{d \psi (r)}{d r}+s_4(r)\psi (r) = 0.
\end{eqnarray}
This equation, (including now the occurrence of the exponential factor $e^{-\eta r}$), has the same shape as \eqref{sch1}, and therefore the same conclusion with \eqref{malade} will be made.

\section{Hydrogen atom in noncommutative space with compactified extra dimension}\label{sec3}
In this section we consider $(D+1)$ NC spacetime , where one dimension $x^D$ is compactified on a circle of radius $R$. This means that $\mathbb{R}^{D+1}$ is reduced to $\mathbb{R}^{D-1+1}\times [-\pi R,\pi R]$ and $x^D=\omega-2\pi nR$, $n\in\mathbb{Z}$. The interaction potential \eqref{intpot} written  now with the required coordinates $\vec r= (r, \alpha_1,\cdots, \alpha_{D-2}) $ and the compactify coordinate $w$ is
\begin{eqnarray}
V(\vec r, w) &=& -q_e^2\sum_{n=-\infty}^{\infty}\Bigg\{\frac{1}{\big(r^2 + (w-2\pi n R)^{2}\big)^{\frac{D-2}{2}}} \cr &+& \frac{i(D-2) r\theta}{\big(r^2 + (w-2\pi n R)^{2}\big)^{\frac{D}{2}}} \partial_0\Bigg\},
\end{eqnarray}
where $ r$ is now the radial coordinates in $(D-1)$-dimensional space, and the  extra dimension  $x^D$ satisfy the condition $\vert x^D -2\pi n R\vert \le \pi R$.  For $D=4$, we get
\begin{eqnarray}\label{pot}
V(\vec r,w) &=&  -q_e^2 \sum_{n \in \mbZ}\Big\{\frac{1}{r^2 + (w-2\pi n R)^{2}} +\cr 
&& \frac{2ir \theta}{\big(r^2 + (w-2\pi n R)^{2}\big)^{2}} \partial_0\Big\}.
\end{eqnarray}
Then  we can compute the followings identities:
\begin{eqnarray}\label{newphan}
\sum_{n \in \mbZ}\frac{1}{r^2 + (w-2\pi n R)^{2}} = \frac{1}{2R r}\frac{\sinh(r/R)}{\cosh(r/R)-\cos(w/R)},
\end{eqnarray}
and
\begin{eqnarray}
\sum_{n \in \mbZ}\frac{2 r}{\big(r^2 + (w-2\pi n R)^{2}\big)^2} =G(r) +F(r)
\end{eqnarray} 
where
\begin{eqnarray}
&&G(r)= -\frac{1}{2R r^2}\frac{\sinh(r/R)}{\cosh(r/R)-\cos(w/R)} \\
&&F(r)= \frac{1}{2R^2 r}\frac{1-\cosh(r/R) \cos(w/R)}{\big(\cosh(r/R)-\cos(w/R)\big)^2}.
\end{eqnarray}
The potential $V(\vec r,w)$ is periodic with  respect to the $w$-direction, and  it can be expanded to a Fourier series as
\begin{eqnarray}
V(\vec r,w) = \sum_{n \in \mbZ} a_n(r)e^{inw/R} +
i\theta\,\sum_{n \in \mbZ} b_n(r)e^{inw/R} \partial_0,
\end{eqnarray}
where
\begin{eqnarray}
&& a_n(r) = -\frac{q_e^2}{2rR}e^{-|n|r/R},\cr
&& b_n(r) =  -\frac{q_e^2}{2Rr^2}\Big[ 1+ |n|\Big]e^{-|n|r/R},
\end{eqnarray}
and such that 
\begin{eqnarray}\label{V_{comp}}
 &&V(\vec  r,w) = \cr
&&-\frac{q_e^2}{2rR} \sum_{n \in \mbZ} \Big[1 + i\theta \Big(\frac{1}{r} + \frac{|n|}{R}\Big)\partial_0 \Big]e^{-(|n|r - inw)/R}.
\end{eqnarray}
The separation of the  variables in the Schr\"odinger equation shows that the radial function $\psi(r)$ satisfies

\begin{eqnarray}
\Big(\frac{d^2}{d r^2} + \frac{2}{r}\frac{d}{d r}- \frac{n^2}{R^2} + \alpha^2+ \frac{\zeta}{r} - \frac{\nu^2_\theta}{r^2}\Big) \psi_n(r) = 0 , 
\end{eqnarray}
 with 
$
\alpha^2 = \frac{2m E}{\hbar^2},\, \nu_\theta^2 = \lambda_4 - \zeta E \theta/\hbar ,\,\zeta =  \nu^2/(2R),\,\nu^2 = \frac{2m q_e^2}{\hbar^2}.
$

The solution of the above equation 
 is  expressed as

\begin{eqnarray}\label{wave}
\psi_{n}(r) &=& \Big[ \frac{(2 \zeta)^{a+2} \; l!}{(2l+1+a)^{a+3} \Gamma(l+1+a)}\Big]^{\frac{1}{2}} \, r^{\frac{1}{2}(a-1)}e^{-\frac{\zeta r}{2l+a+1}}
\cr 
&& \times L_{l}^{a}\Big(\frac{2 \zeta r}{2l+a+1} \Big), \quad a = \sqrt{4 \nu_\theta ^2+1}, 
\end{eqnarray}
where we have used the normalisation condition $
\int_0^{\infty}e^{-z} z^{a+1} [L_{l}^{a}(z)]^2 dz = \frac{(2l+1+a) \Gamma(l+1+a)}{l!}
$, and 
where $L_l^a$ stands for the generalized Laguerre polynomial. The quantum number  $l$ is  a positive integer, which correspond to the physical situation. This integer  is  given by
\begin{eqnarray}\label{zaaa}
l = -\frac{1}{2}-\frac{1}{2}\sqrt{4 \nu_\theta^2+1}+\frac{\zeta  R}{2\sqrt{n^2-\alpha^2 R^2}}.
\end{eqnarray}
Two energies contribution appear from the relation \eqref{zaaa}.
\bea\label{e1}
E_{nl}^{(1)}=\frac{\hbar^2}{2m}\Big( \frac{n^2}{R^2} - \frac{\zeta^2}{(2l + 1 +\sqrt{1+4\lambda_4})^2}\Big)
\eea
and
\bea\label{e2}
E_{nl}^{(2)}&=&\frac{\hbar^2}{2m}\frac{\zeta^2}{(2l+1+\sqrt{1+4\lambda_4})^2}\cr
&+&\frac{\hbar\sqrt{1+4\lambda_4}}{4\theta\zeta}(2l+1+\sqrt{1+4\lambda_4}).
\eea
Let us discuss the energy spectrum  \eqref{e2}. In the limit where $\theta\rightarrow 0$,  $E_{nl}^{(2)}$ is not well defined. Also as expected in our introduction the eigenfunctions of the operator $\nabla^2_{x^4}$ with eigenvalues $E_{n0}=-n^2/R^2$ is not recovered. Finally this expression can not be taking into account as solution of eigenvalues problem. Then  the energy spectrum becomes
\begin{eqnarray}\label{solenergy1}
E_{nl} = \frac{\hbar^2}{2m}\Big( \frac{n^2}{R^2} - \frac{\zeta^2}{(2l + 1 +\sqrt{1+4\lambda_4})^2}\Big).
\end{eqnarray}
\begin{remark}
\begin{itemize}
\item Our  result shows that
the energy spectrum \eqref{solenergy1}  does not depend on the NC parameter $\theta$  if we consider the first order approximation of this parameter. The solution of eigenvalue problem of the hydrogen atom with compactified one dimension is solved numerically in \cite{Bures:2015hfa}-\cite{Bures:2014fza} (see also \cite{Chaichian:2002ew}-\cite{Chaichian:2000si} in the case where no dimensions are compactified).  Due to the fact that $\lim_{\theta\to 0} E_{nl}=E_{nl}$, the expression \eqref{solenergy1} can be condered as the solution of Hydrogen atom  in $4+1$-dimensional spacetime for both NC\footnote{  This energy is valid where the first order approximation in $\theta$ is considered} and commutative case, where one dimension is compactified.
\item
 The quantity 
\bea
E_{nl}'=-\frac{\hbar^2\zeta^2}{2m(2l + 1 +\sqrt{1+4\lambda_4})^2}.
\eea 
 correspond to the reduce dimension energy spectrum and is discussed in the introduction  of our paper.

\end{itemize}
\end{remark}
We consider now the case of Yukawa potential \eqref{Yukawa}  for $D = 4$. On shell, and compactified the $x^4$  direction on the circle we get the reduce potential
\begin{eqnarray}\label{relationm1}
V(\vec r,w) &=&i V_0 \theta\, \Big[-\frac{1}{2R r^2}\frac{\sinh(r/R)}{\cosh(r/R)-\cos(w/R)} \cr
&+& \frac{1}{2R^2 r}\frac{1-\cosh(r/R) \cos(w/R)}{\Big(\cosh(r/R)-\cos(w/R)\Big)^2}\Big] \partial_0 \cr
&-&\frac{V_0}{2R r}\frac{\sinh(r/R)}{\cosh(r/R)-\cos(w/R)}.
\end{eqnarray}
Let us briefly give the proof of this relation. The goal of this prove is to compute the integral $\int_{\Gamma}\, f(z)dz$, where $\Gamma$ is a closed contour on the complex plane and $f(z)$ is a  holomorphic function given by
\begin{eqnarray}
 f(z) = \frac{\cot(\pi z)\exp\big[-c\big((a-z)^2 + b^2\big)^{1/2}\big]}{(a-z)^2 + b^2}, 
\end{eqnarray}
where $a,b,c$ are three real  numbers.  The pole of  $f(z)$ are $z_n = n,  n \in \mbZ$, $ z_I = a+ib $ and $z_{II} = a-ib$. Using the residue theorem:
\beq
\int_{\Gamma}\, f(z)dz=2i\pi \sum \mathcal{R}es [f(z)]=0,
\eeq
with
\begin{eqnarray}
&&\underset{z_n \quad \quad}{\mathcal{R}es[f(z)]} = \frac{\exp\big[-c\big((a-n)^2 + b^2\big)^{\frac{1}{2}}\big]}{\pi [(a-n)^2 + b^2]},\cr 
&&\underset{z_I \quad \quad}{\mathcal{R}es[f(z)]} = -\frac{i}{2b}\cot[\pi(a+i b)] , \quad \mbox{ and} \cr
&&  
\underset{z_{II} \quad \quad}{\mathcal{R}es[f(z)]} = \frac{i}{2 b}\cot[\pi(a-i b)]
\end{eqnarray}
 Hence,
\bea
 &&\sum_{n \in \mbZ}\frac{\exp\big[-c\big((a-n)^2 + b^2\big)^{1/2}\big]}{(a-n)^2 + b^2} \cr
&&= \frac{\pi}{b} \frac{\sinh(2\pi b)}{\cosh(2\pi b) - \cos(2\pi a)}.
\eea
The first term on the right hand side of \eqref{relationm1}is 
\begin{eqnarray}\label{nouv}
&& \sum_{n \in \mbZ}\frac{\exp\big[-\zeta \big(r^2 +(w -2\pi n R)^2 \big)^{1/2}\big]}{r^2 + (w-2\pi n R)^2} 
 \cr
&&= \frac{1}{2 R r} \frac{\sinh(r/R)}{\cosh(r/R) - \cos(w/R)}.
\end{eqnarray}
The second term on the right hand side of expression \eqref{relationm1} is the first order derivative of equation \eqref{nouv} respect to $r$. Then the  relation \eqref{relationm1} is straightforward obtained.

Now using the fact that the fonction $V(\vec r ,w)$ is periodic,  the Fourier serie can be given by
\begin{eqnarray}
V(\vec r,w) = \sum_{n \in \mbZ} \Big(a_n(r) +
i\theta b_n(r) \partial_0\Big)e^{inw/R},
\end{eqnarray}
where the Fourier coefficients  are:
\begin{eqnarray}
a_n(r) &=& -\frac{V_0}{2rR}e^{-|n|r/R}, \cr
b_n(r) &=& -V_0 \Big[\frac{1}{2Rr^2} + \frac{|n|}{2R^2 r}\Big]e^{-|n|r/R}
\end{eqnarray}
Finally we come to
\begin{eqnarray}
V(\vec r,w) = &-&\frac{V_0}{2rR} \sum_{n \in \mbZ} \Big[1 + i\theta\Big(\frac{1}{r} + \frac{|n|}{R}\Big)\partial_0 \Big]\cr
&\times&e^{-(|n|r - inw)/R}.
\end{eqnarray}

the relation \eqref{relationm1} can also be expand using the Fourier serie as
\begin{eqnarray}
V(\vec r,w) = &-&\frac{V_0}{2rR} \sum_{n \in \mbZ} \Big[1 + i\theta\Big(\frac{1}{r} +  \frac{|n|}{R}\Big)\partial_0 \Big]\cr
&\times&e^{-(|n|r - inw)/R}
\end{eqnarray}
and  the radial equation takes the form
\begin{eqnarray}
\Big[\frac{d^2}{d r^2} + \frac{2}{r}\frac{d}{d r}- \frac{n^2}{R^2} + \alpha^2+ \frac{u}{r} - \frac{v}{r^2}\Big] \psi_n(r) = 0,
\end{eqnarray}
 where  $\alpha^2 = \frac{2m E}{\hbar^2}$;  $v = \lambda_4 -\frac{m V_0 E \theta}{\hbar^3 R}$, $u = \frac{m V_0}{\hbar^2 R}$. The solution of this equation leads to the same results given in \eqref{wave} and \eqref{solenergy1}.

\section{Discussion and conclusion}\label{sec4}
In this paper we have found that the noncommutativity of spacetime can help to compute the exact  expression of the energy spectrum of the hydrogen atom in  $(4+1)$ dimensions with compactified one extra dimension. Unfortunately, it's clear that this method can not be used in higher dimensions. To be more precise, let us consider
the particular case where $D=6$, the compactified one dimension $x^6$ gives the potential
\bea\label{show}
V(\vec r,w)&=& -\frac{q_e^2}{(2 R r)^{2}} \Big(\frac{R}{r} \frac{\sinh(r/R)}{\cosh(r/R) - \cos(w/R)} \cr
&+& \frac{\cosh(r/R) \cos(w/R) - 1}{[\cosh(r/R) - \cos(w/R)]^2} \Big) \cr
&+&  \frac{i \theta\,q_e^2}{8 r^4 R^3 \big(\cos(w/R)-\cosh(r/R)\big)^3}
\cr
&\times&\Big[
(r^2+3 R^2) \sinh(r/R) \cos(2w/R) \cr
&+& \cos(w/R) \Big((r^2-6 R^2) \sinh(2r/R) \cr
&+& 3 r R (\cosh(2r/R)+3)\Big)\cr
&+& 3 \sinh(r/R) \Big(R^2 \cosh(2r/R)+2 R^2 -r^2\Big) \cr
&-& 3 r R \cosh(r/R) \Big(\cos(2w/R)+3\Big)\Big]\partial_0.
\eea
This relation can be expanded  in a Fourier series as
\begin{eqnarray}\label{V_{comp}}
&&V(\vec r,w) = -\frac{q_e^2}{(2 R r)^{2}}\sum_{n \in \mbZ} \Big[\Big(\frac{R}{r} + |n| \Big)\cr
&&+i\theta\, \Big(\frac{3R}{r^2} + 
  \frac{3|n|}{r} + \frac{n^2}{R} \Big) \partial_0\Big]  e^{-|n|r/R} e^{inw/R},
\end{eqnarray}
in which the radial part of the Schr\"odinger equation become 
\begin{eqnarray}
&\Big(\frac{d^2}{d r^2} + \frac{2}{r}\frac{d}{d r}-\frac{\lambda_4}{r^2} - \frac{n^2}{R^2} + \alpha^2 + \frac{2m q_e^2}{4\hbar^2r^3 R} + \frac{3m q_e^2E \theta}{2\hbar^3  r^4 R} \Big) \psi_n(r) \cr
&=0.
\end{eqnarray}
The solution of this equation is not yet understood. Surprisingly, we have also show that, despite from this noncommutativity, the energy spectrum do not depend on the deformation parameter $ \theta$ and therefore might be considered as the energy solution of commutative space,  with  compactified one extra dimension, solved in \cite{Bures:2015hfa} and \cite{Bures:2014fza}. Finally let us mention that in the case of higher dimensions more than $4+1$ the compactified several extra dimensions may be considered.

Let us examinated the case of the  Klein-Gordon (KG) equation:
\begin{equation}\label{KG}
\begin{split}
 &\Big(i\hbar\partial_0 - V(\vec{r}_{nc})\Big)^2 \psi(\vec{r}_{nc}, t)  \cr
 &= m^2 c^4 \psi(\vec{r}_{nc}, t)- \hbar^2 c^2 \Delta \psi(\vec{r}_{nc}, t),
\end{split}
\end{equation}
where the potential $V(\vec{r}_{nc})$ is 
\beq
V(\vec{r}_{nc})=-\wp \frac{\hbar c}{|\vec r_{nc}|^{D-2}}.
\eeq
$\wp$ is related to the fine structure constant and $c$ is the light speed. For $D=4$,
 by taking into account the fact that $\vec{r}_{nc}=\vec{r}-i\vec\theta\partial_0$,  and writing  the extra-dimension $x^4=w - 2\pi R k,\,\, k\in\mathbb{Z}$,  we get
\begin{eqnarray}\label{wao}
V(r, w) &=& -\wp \hbar c \sum_{n=-\infty}^{\infty}\Big\{\frac{1}{r^2 + (w-2\pi n R)^{2}} \cr &+& \frac{2i r\theta}{\big(r^2 + (w-2\pi n R)^{2}\big)^{2}} \partial_0\Big\}.
\end{eqnarray}
  The left hand side of the equation \eqref{KG}, using \eqref{wao} gives 
\begin{eqnarray}\label{oui}
&& \Big(i\hbar \frac{\partial}{\partial t} - V({r}, w)\Big)^2\psi(\vec r,t) = \cr
&&\Big[  -\hbar^2 \partial_0^2  +  \wp \hbar^2 c \Big(2i A (r,w) \partial_0  + 3\theta  \frac{\partial A(r,w)}{\partial r} \partial_0^2\Big) \cr
&& +   \wp^2 \hbar^2 c^2 \Big( A^2(r,w)  -  i\theta \frac{\partial A^2(r,w)}{\partial r} \partial_0\Big)\Big]\psi(\vec r,t)
\end{eqnarray}
where $A(r,w)$ is  written  as
\begin{eqnarray}
A (r,w) = \frac{1}{2 R r} \frac{\sinh(r/R)}{\cosh(r/R) - \cos(w/R)},
\end{eqnarray}
with  the Fourier serie
\begin{eqnarray}\label{malfra1}
A (r,w)  = \frac{1}{2rR} \sum_{n \in \mbZ}  e^{inw/R}  e^{inw/R}.
\end{eqnarray}
After some technical handling, we can  show that the  quantity $A^2(r,w)$ is expanded  as
\begin{eqnarray}\label{malfra2}
A^2(r,w)  =  \frac{1}{(2 R r)^2} \sum_{n \in \mbZ} \Big[|n| + \coth\Big(\frac{r}{R}\Big) \Big] e^{-\frac{|n|r}{R}} e^{\frac{inw}{R}}.
\end{eqnarray}
By replacing \eqref{malfra1} and \eqref{malfra2} in  \eqref{oui} and 
separating variables as $\psi(\vec{r}, t) = \Psi(\vec{r}) e^{- i E t/\hbar} $ and  $\Psi(\vec{r}) = \Psi(r) \mathcal{Y}_\ell^{(3)}$ such that $\hat{l}^2 \mathcal{Y_\ell}^{(3)} = \ell(\ell + 2)\mathcal{Y_\ell}^{(3)} =\lambda_4 \mathcal{Y_\ell}^{(3)} $ we find the radial equation: 
\begin{eqnarray} \label{eqnaprox}
&\Big[\frac{1}{r^2}\frac{d}{dr}\Big(r^2\frac{d}{dr}\Big)+V_{\rm eff}(r) + \frac{E^2}{\hbar^2 c^2}  - \frac{m^2 c^2}{\hbar^2} - \frac{n^2}{R^2}\Big] \Psi_{n}(r)  = 0,\cr
&
\end{eqnarray}
where the effective potential  $V_{\rm eff}(r)$ is 
\bea
V_{\rm eff}(r)&=&-\frac{\lambda_4}{r^2} + \frac{3\theta \wp E^2}{2r^2R\hbar^2 c} +  \frac{\wp E}{rR\hbar c}+ \frac{\wp^2 }{(2 R r)^2} \coth(r/R)\cr
&& +   \frac{\theta \wp^2  E}{(2 R r)^2 \hbar} \Bigg(\frac{2}{r}\coth (r/R) + \frac{ r}{\sinh^2(r/R)}\Bigg) 
\eea
\begin{figure}
\begin{center}
\includegraphics[scale=0.35]{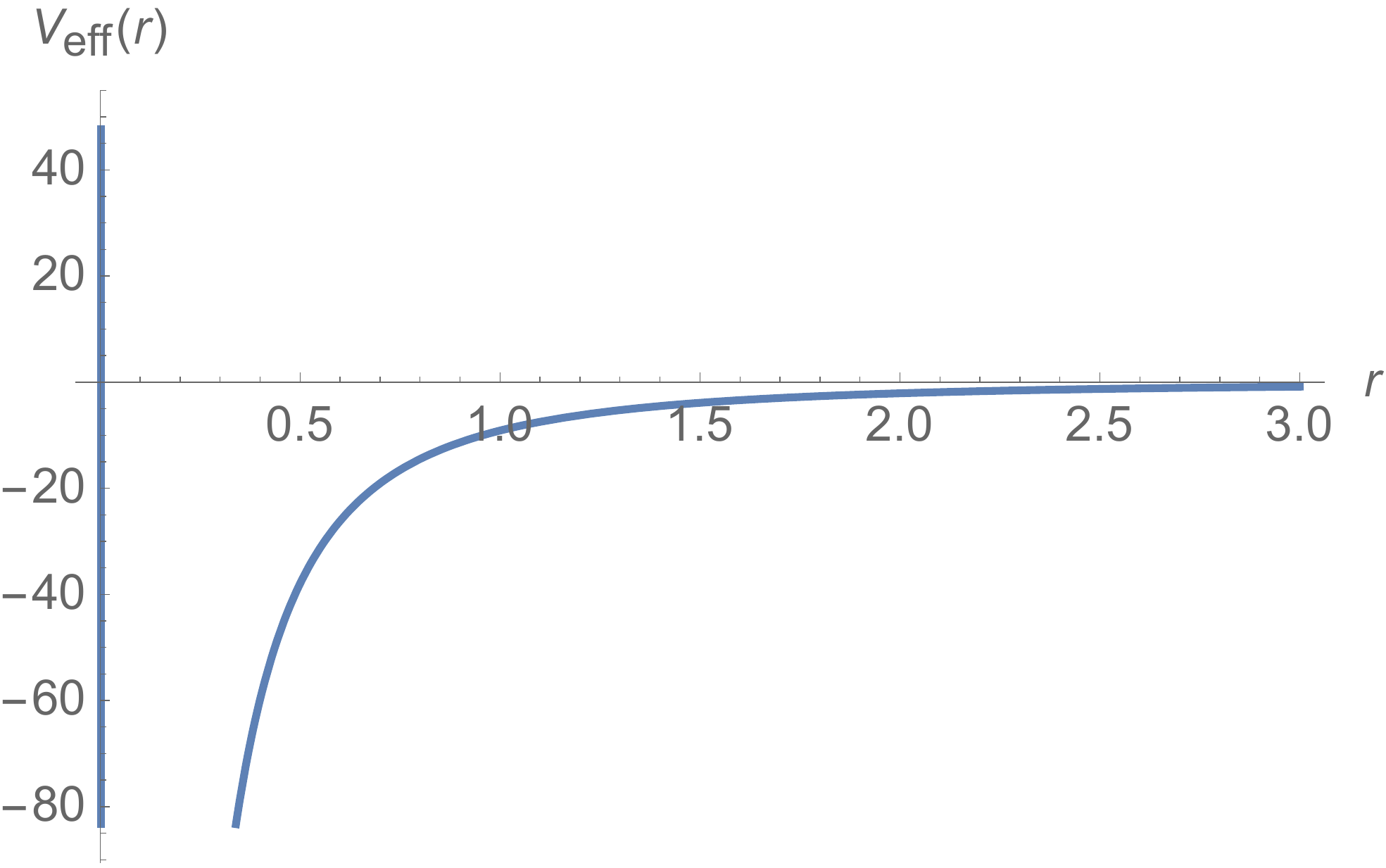}
\end{center}
\caption{Plot of the potential $V_{\rm eff}(r)$, with $R=0.01$, $\theta=0.01$, $\hbar=c=1,$  $\wp=1/137$, $E=1$, $\ell=1$.}\label{fignew}
\end{figure}
The  equation  \eqref{eqnaprox} can be solved numerically using the approximation method.  Consider the Taylor expansion of $V_{\rm eff}$  by using the fact that: 
\begin{eqnarray}
&&  \coth(r/R) =  \frac{R}{r} + \frac{r}{3R} + ... \cr
&& \sinh(r/R) = \frac{r}{R}  + ...
\end{eqnarray}
Then \eqref{eqnaprox} becomes
\begin{eqnarray}\label{svgarde}
&&\frac{d^2 \Psi_{n}(r)}{d r^2} + \frac{2}{r}\frac{d \Psi_{n}(r)}{d r}+\sum_{j=0}^4\,\frac{b_j}{r^j} \Psi_{n}(r)  = 0. \;\;\;
\end{eqnarray}
with
$
 b_4 = \frac{\theta \wp^2  E}{2 R \hbar},  \,\,
b_3= \frac{\theta \wp^2  E}{4 \hbar} +  \frac{\wp^2 }{4 R},\,\,
 b_2 = \frac{\theta \wp^2  E}{6 R^3 \hbar} + \frac{3\theta \wp E^2}{2 R\hbar^2 c} - \lambda_4, \,\,
 b_1 = \frac{\wp^2 }{12 R^3} +  \frac{\wp E}{R\hbar c} ,\,\,
 b_0 = \frac{E^2}{\hbar^2 c^2}  - \frac{m^2 c^2}{\hbar^2} - \frac{n^2}{R^2}.
$

We first examinated  the wave functions $\Psi_n(r)$ in the asymptotic range $r\rightarrow \infty$. The 
  potential $V_{\rm eff}(r)$ vanish, in this limit, i.e.
\bea
V_{\rm eff}(r_\infty)\rightarrow 0.
\eea
In the region $r_\infty$ the equation \eqref{svgarde} gives the solution of the form 
\bea
\Psi^\infty_n(r)=U\frac{e^{-\sqrt{-b_0}r}}{r},\quad U\in \mathbb{R}.
\eea
The general solution of the equation \eqref{svgarde} takes the form
\bea
\Psi_n(r)=U(r)\frac{e^{-\sqrt{-b_0}r}}{r},
\eea
where $U(r)$ satisfy the differential equation
\bea
U''(r)-2\sqrt{-b_0}U'(r)+\sum_{j=1}^4\,\frac{b_j}{r^j} U(r)  = 0
\eea
The investigation of the numerical solution of this equation can be made in forthcoming work.
\section{Acknowledgments}
DOS research at Max-Planck Institute  is supported by the Alexander von Humboldt foundation. The authors are grateful to the referee for his useful comments that allowed to improve the
paper.

\end{document}